\documentclass[aip,rsi,graphicx,groupedaddress,reprint]{revtex4-1} % for checking your page length
\usepackage[colorlinks=true,urlcolor=blue,citecolor=blue,linkcolor=blue]{hyperref}
\usepackage[T1]{fontenc}
\usepackage[latin9]{inputenc}
\usepackage{times}
\usepackage{color}
\usepackage{amssymb,amsmath}
\usepackage{amsbsy}
\usepackage{graphicx}
  {\addtolength{\leftskip}{#1}\addtolength{\rightskip}{#2}}{\par}

\begin{document}

\title{\large Note: Computer controlled rotation mount for large diameter optics}
\author{Ana Rakonjac}
\affiliation{Jack Dodd Centre for Quantum Technology, Department of Physics, University of Otago, Dunedin, New Zealand.}
\author{Kris O. Roberts}
\affiliation{Jack Dodd Centre for Quantum Technology, Department of Physics, University of Otago, Dunedin, New Zealand.}
\author{Amita B. Deb}
\affiliation{Jack Dodd Centre for Quantum Technology, Department of Physics, University of Otago, Dunedin, New Zealand.}
\author{Niels Kj{\ae}rgaard}\email{niels.kjaergaard@otago.ac.nz}
\affiliation{Jack Dodd Centre for Quantum Technology, Department of Physics, University of Otago, Dunedin, New Zealand.}
\date{\today}

\begin{abstract}
%\begin{custommargins}{0cm}{2cm}
%  \noindent\leftarrowfill Same left margin. Smaller right margin\rightarrowfill ill
%\end{custommargins}{
We describe the construction of a motorized optical rotation mount with a 40~mm clear aperture. The device is used to remotely control the power of large diameter laser beams for a magneto-optical trap (MOT). A piezo-electric ultrasonic motor on a printed circuit board (PCB) provides rotation with a precision better than 0.03$^\circ$ and allows for a very compact design. The rotation unit is controlled from a computer via serial communication, making integration into most software control platforms straightforward.
\end{abstract}
%\pacs{34.50.-s, 05.30.Jp, 31.15.xv}
\maketitle
\begin{figure*}[t!]
\begin{center}
    \includegraphics[width=\textwidth]{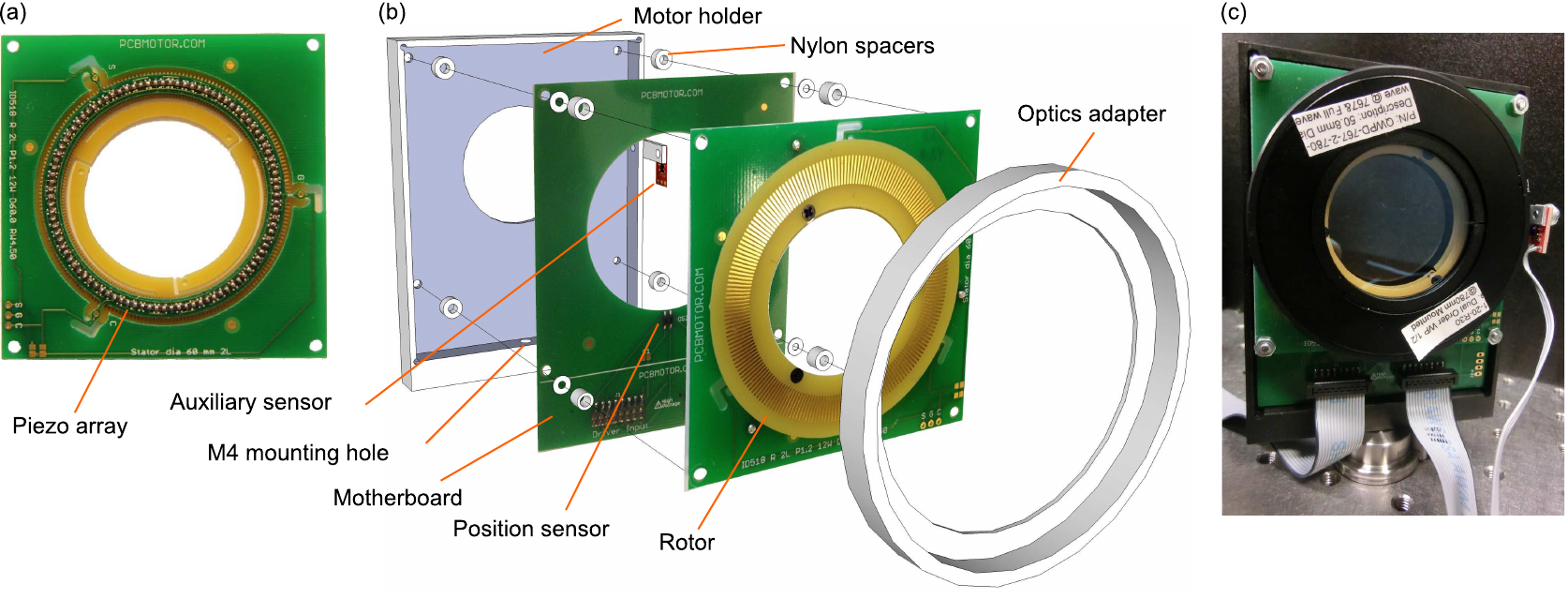}
\caption{(Color online) (a) The PCB stator. (b) The exploded rotator motor with rotor mounted. Ref. \onlinecite{supp} provides mechanical drawings for the\label{fig:motor} motor holder and the optics adapter. (c) The assembled rotator; video showing rotator operation (enhanced online)[\href{www.physics.otago.ac.nz/staff_files/nk/files/movie1.mp4}{URL: www.physics.otago.ac.nz/staff\_files/nk/files/movie1.mp4}].
.\label{fig:motor}}
\end{center}
\end{figure*}
Controlling the polarization state of light via rotating elements such as half- and quarter-wave retarders and polarizers is extensively used in optical instrumentation for purposes such as polarimetry and ellipsometry \cite{Clarke1971, Arwin2009}. In instruments where the polarization state is to be scanned, stepper motors or DC servo motors are typically employed \cite{Fueten1997,El-Agez2011}. These motors are almost invariably not an integral part of the rotation mount holding the optics. Rather, they are mounted externally and exert a torque on the rotation mount via a gearing mechanism. Hence, such rotators are typically not very compact. A recipe for constructing a relatively small rotator for a polarizer was recently provided in the present journal\cite{shelton2011}. This was accomplished by replacing the rotor axle of a stepper motor with a lens tube resulting in a direct drive. The clear aperture for this rotator was, however, limited to 10~mm, while the outer diameter of the motor was 42~mm. Compact rotators based on piezo motors are commercially available and provide an alternative solution to stepper and DC motors. However, to our knowledge, the maximum clear aperture is limited to one inch for existing devices of this type.

In this Note, we describe an implementation of a compact \textit{direct drive} rotator for optical components with a 40 mm clear aperture. The development of this motorized rotator is specifically motivated by a need in our laboratory to remotely control the angle of the fast axis of large diameter half wave plates as part of the optical setup for a dual atomic species MOT\cite{Goldwin2002}. These half-wave plates are to be mounted in a constrained space, where lack of access precludes manual adjustment. With commercially available rotators either too bulky to fit or having a limited clear aperture, a solution based on a piezo-electric ultrasonic motor kit was developed as described below.

We base our rotator on an ultrasonic motor evaluation kit from PCBMotor\cite{Note1}, which comes complete with a motor driver and a power supply unit (PSU). The driver interfaces to a computer via a USB connection acting as a virtual COM port and can control up to two motors (more can be added through multiplexing). Figure ~\hyperref[fig:motor]{\ref*{fig:motor}(a)} shows the heart of the motor -- the 60~mm stator -- which is a PCB (glass-reinforced epoxy laminate) with a circular array of piezo elements on both sides mounted on an annulus suspended at three pivot points. The piezo elements are driven by two oscillating voltage sources that each excite the array with a standing wave. The excitations are applied such that these two standing waves are 90$^\circ$ out of phase in both space and time. This results in a traveling wave around the circular array in either the clockwise or counter-clockwise direction, depending on the time-space phase relation\cite{Uchino1998}. Figure~\hyperref[fig:motor]{\ref*{fig:motor}(b)} shows the motor with the spring-loaded rotor in place. The motor (stator and rotor) comes mounted on an $80\times100$~mm motherboard via four corner holes in the stator. An infrared emitter/sensor on this motherboard monitors rotation via 200 reflective copper encoder lines on the rotor.
Due to the frictional coupling mechanism between the rotor and the stator, the motor provides a holding torque even in an unpowered or disconnected state. We measured this holding torque to be 0.066~Nm. When driven at the rated maximum voltage of 5~V, we found the motor capable of delivering a stall torque of 0.052~Nm and rotating at angular speed of  $\rm 155^\circ /s$.

To modify the ultrasonic motor into an optical rotator, we replace the original four M3 corner screws with longer ones (16~mm) so that the motor unit can be mounted inside a milled out aluminum holder on 3~mm long cylindrical nylon spacers. The aluminum holder is anodized black to reduce stray light reflections and has an M4 threaded hole for screwing onto a standard optical post. A cylindrical optics adapter mount is glued directly onto the rotor using epoxy resin. A small stand-off rim on this adapter ensures that the rotor is centered and improves adhesion. An assembled rotator is shown in Fig.~\hyperref[fig:motor]{\ref*{fig:motor}(c)}.

To define an origin for rotation, a small auxiliary line sensor (a breakout board version\cite{Note2} of a Fairchild QRE1113 reflective object sensor\cite{Note3}) is mounted on one side of the motor holder [Fig.~\hyperref[fig:motor]{\ref*{fig:motor}(b)}] to detect a zero point line on the rim of the optics adapter. We read out the sensor value using an ADuC7020 microcontroller. This is available as an inexpensive ``Minikit" development system which comes complete with a serial communication cable and software for programming the microcontroller\cite{Note4}. The microcontroller board is powered from the PCBMotor 5~V PSU and in turn powers the auxiliary sensor board from its 3.3~V AVDD supply. The compiled program code in Ref.~\onlinecite{supp} will, when downloaded to the microcontroller\cite{Note5}, sequentially read the analog inputs on channels 1, 2, 3, and 4 and output text strings with the voltage information to a serial port.

\begin{figure}[b!]
\begin{center}
    \includegraphics[width=\columnwidth]{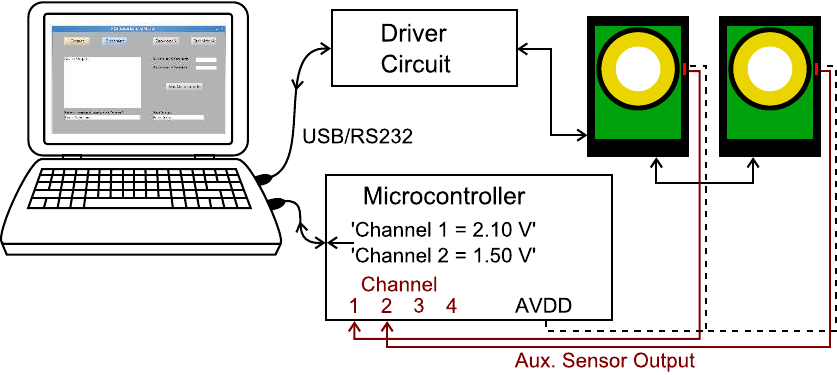}
\caption{(Color online) Outline of the communication scheme. The motor driver circuit and the microcontroller communicate with the control program via virtual USB serial ports. The motor system and microcontroller share the same 5 V power supply. The auxiliary sensors output a voltage directly to the microcontroller.}
\label{fig:blockdiagram}
\end{center}
\end{figure}
\begin{figure}[b!]
\begin{center}
    \includegraphics[width=\columnwidth]{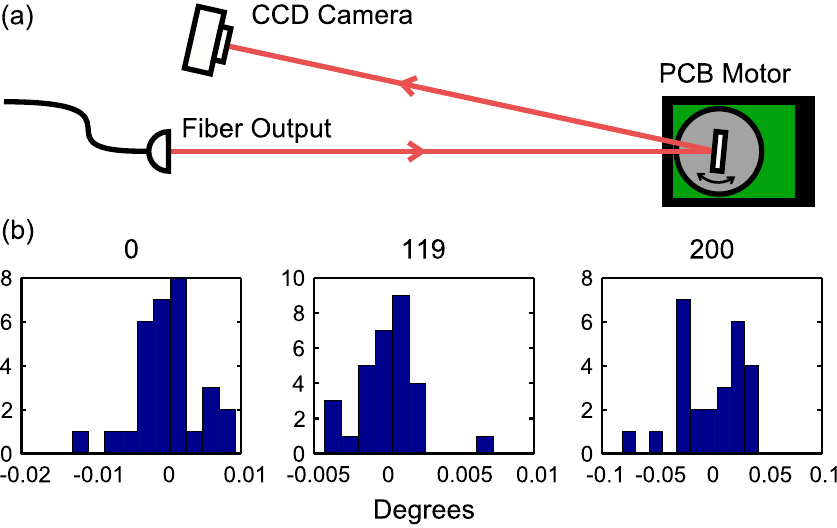}
\caption{(Color online) (a) Setup for measuring position repeatability for the motor. A collimated laser beam is reflected onto a CCD camera by a mirror mounted perpendicular to the motor. Beam position is determined using a Gaussian fit. (b) Histograms of repeatability data for three different points (0, 119, 200) on the code wheel. 0$^\circ$ indicates the mean position for each data set.}
\label{fig:angres}
\end{center}
\end{figure}
\begin{figure}[t!]
\begin{center}
    \includegraphics[width=\columnwidth]{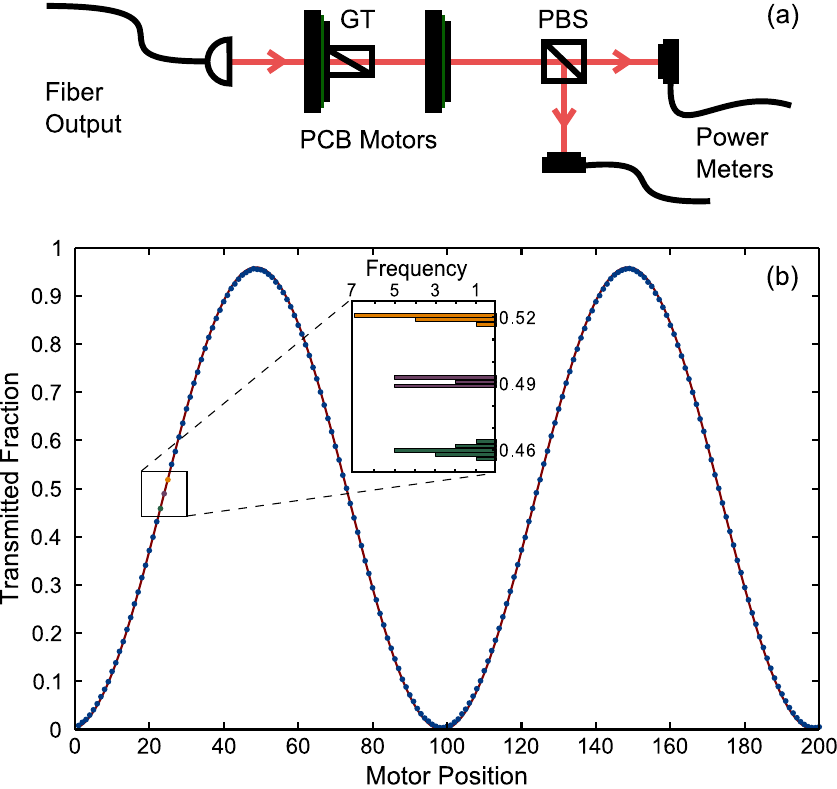}
\caption{(Color online) (a) Setup for investigating power control functionality of the motorized wave plate. A GT polarizer ensures good polarization purity prior to the wave plate and both outputs of the PBS are monitored using a pair of identical power meters. (b) Transmitted fraction of light through a multi-order wave plate ($\lambda/2$ for 767 nm) for 200 steps, corresponding to 180$^\circ$. Data points are averages over 12 consecutive experimental runs. The inset shows histograms for motor positions 23, 24 and 25.}
\label{fig:polfig}
\end{center}
\end{figure}
Since both the motor driver unit and the microcontroller communicate with serial ports on a computer, it is straightforward to write a customizable control program on system design platforms, such as LabVIEW or Matlab. In Ref.~\onlinecite{supp}, the code for a Matlab\cite{Note6} sample program with basic functionality (requires the Instrument Control Toolbox) and a graphical user interface is provided. The communication arrangement is shown in Fig.~\ref{fig:blockdiagram}.

As mentioned above, the rotor has a code wheel with 200 lines. With the sensor on the mother board triggering on both leading and trailing edges of the lines, readback from the encoder divides a full rotation into 400 increments (unit steps). The controller can rotate the motor a given number of unit steps in either direction, but we restrict rotations to the clockwise direction to ensure consistency in step size. In between unit steps, it is possible to achieve very fine steps of 0.005$^\circ$ using so-called micropulses, where small motor movements are executed by applying power to the piezo array in short bursts. The control program locates the \textit{absolute} zero point (origin) via the externally mounted sensor by unit-stepping the rotor until the zero point line is detected.

We characterize the repeatability in positioning the rotator using the setup shown in Fig.~\hyperref[fig:angres]{\ref*{fig:angres}(a)}. By mounting a mirror normal to the axis of rotation, we observe a variation of 0.005$^\circ$ (standard deviation) in the position of a reflected laser beam upon performing the zeroing operation (origin repeatability). In general, however, we found some variation in repeatability for different encoder values with a few giving rise to fluctuations up to 0.03$^\circ$ [see Fig.~\hyperref[fig:angres]{\ref*{fig:angres}(b)}]. Despite this variability, positioning accuracy far exceeds what can be achieved manually and is comparable to commercially available motorized mounts.

As our application of the motorized rotation mounts is to control the power ratio for laser beams transmitted and reflected by polarizing beam splitter (PBS) cubes, we investigated their functionality in this context. A nominal half wave plate was inserted in one of our motorized rotation mounts and placed between a Glan-Thompson (GT) polarizer and a PBS cube in a crossed orientation [see Fig.~\hyperref[fig:polfig]{\ref*{fig:polfig}(a)}]. Prior to inserting the waveplate in the beam path, the crossed configuration was obtained by rotating the GT polarizer --  we used the second motorized rotation mount for this purpose -- and minimizing the transmission of the beam through the PBS as measured using an optical power meter. Figure~\hyperref[fig:polfig]{\ref*{fig:polfig}(b)} shows the transmitted fraction of light\cite{Collett1993} $\propto (1-\cos4\theta)$ with the waveplate inserted and rotated over an angle $\theta=0^\circ - 180^\circ$.
The standard deviation of the transmitted fraction for different points in the inset is characterised by an average of 0.002, indicating excellent control over the polarization. It is possible to access higher resolution than a single step size by employing the micropulse feature of the motor. Though position control is good, the mechanical stability of the device is inherently worse than that of a high quality manual rotation mount due to the PCB construction material, resulting in a measured wobble of 0.74$^\circ$ for the device under consideration. This has a negligible effect on the waveplate retardance\cite{Williams1999}, though it may not be ideal for other applications.

In conclusion, we have described the construction of a large aperture compact motorized rotation mount. Our particular application was polarization control of large radius laser beams for a MOT, but our solution lends itself to a host of optical instruments, e.g., imaging polarimeters \cite{Greaves2003,Pezzaniti1995,Wijngaarden2001}. We finally note that our implementation incorporates a microcontroller with four analog inputs and that only two of those are employed for the positioning (one for each motor). Hence, the control program can easily be extended to acquire data from, for example, photo-detector signals of an apparatus. In fact, this was how we acquired the data for Fig.~\hyperref[fig:polfig]{\ref*{fig:polfig}(b)}.

We thank Callum McKenzie for help on programming the microcontroller.
This work was supported by FRST contract NERF-UOOX0703 and the Marsden Fund of New Zealand
(Contract No. UOO1121).

\bibliographystyle{apsrev4-1}

%\bibliography{rotbibnew}
%

\end{document}